\documentclass[aps,pra,reprint,groupedaddress,showpacs]{revtex4-1}

\usepackage{braket}
\usepackage{makeidx}
\usepackage{multirow}
\usepackage[dvipsnames,svgnames,table]{xcolor}
\usepackage{graphicx}
\usepackage{ulem}
\usepackage{hyperref}
\usepackage{amsmath}
\usepackage{amssymb}

\begin{document}


\title{Phase memory across two single photon interferometers including wavelength conversion}


\author{A. Heuer, S. Raabe, R. Menzel}
\affiliation{Photonik, Institut f\"ur Physik und Astronomie, Universit\"at Potsdam,  D-14469 Potsdam, Germany}

\date{\today}

\begin{abstract}

Spontaneous parametric down conversion (SPDC) in a nonlinear crystal generates two single photons (signal and idler) with random phases, each. Thus, no 1$^{st}$-order interference between them occurs. However, coherence can be induced in a cascaded setup of two crystals, if e.g. the idler modes of both crystals are aligned to be indistinguishable. Due to the effect of phase memory it is found that the first order interference of the signal beams can be controlled by the phase delay between the pump beams.
And even for pump photon delays much larger than the coherence length of the SPDC-photons the visibility is above 90$\%$. These high visibilities reported here prove for an almost perfect phase memory effect across the two interferometers for the pump and the signal photon modes. 
\end{abstract}


\maketitle


\section{Introduction}
It was shown by Grangier et al.\cite{grangier88} that in spontaneous parametric down conversion (SPDC) the generated photon pair en bloc carries the phase information about the pump field although the single photon phases are random. This is the basis of several kinds of two photon interference effects \cite{Jha08, Ou07}. Later Ou et al. \cite{Ou90} described this as phase memory. They demonstrated that SPDC-quantum states can be described as a linear superposition of a two-photon state with vacuum \cite{Ou89}. The light then carries phase information of the involved electromagnetic fields of the light. (Although otherwise a group of photons in a Fock state does not carry phase information.) In their example a coherent pump field interacts with the nonlinear medium to generate photon pairs via SPDC. The down-converted signal and idler photons have no fixed phase relation and therefore no  1$^{st}$-order interference effects between them can occur. But the two photons as a combined entity carry phase information about the pump field which can be observed by biphoton interference experiments \cite{Ou90}. The biphoton phase is determined by the phase of the pump field as described in \cite{Ou89} and here below.\\
The same group demonstrated later \cite{Zhou91} the effect of induced-coherence using again two separated SPDC-crystals pumped by the same coherent pump laser light \cite{Wang91}. But in this experiment the idler output of the first crystal was used to seed the SPDC process in the second crystal generating a new pair of signal and idler photons. In this experiment the photon density was so low that during the measurement interval usually just one photon pair in one of the two crystals was generated. This way these two photon pairs were entangled. As a consequence the two signal photon channels from the two crystals could be contained in a balanced interferometer and single photon ( 1$^{st}$-order) interference at the superposition of the two signal channels was observed  \cite{Zhou91, Wang91, Zhou93, Grayson94a, Grayson94b}.
This experiment can be viewed as a Mach-Zehnder interferometer for a single photon. But because the signal photon can be generated in the  1$^{st}$ or the 2$^{nd}$-crystal the propagation of the phase information across the whole interferometer including both crystals is the precondition for the observation of  1$^{st}$-order interference with the single photon. This was demonstrated, explicitly, by attenuating the idler channel between the first and second crystal using a transmission filter. A decrease of the visibility of the  1$^{st}$-order interference of the signal photon was observed and modeled  \cite{Wang91, Wiseman00}.\\
Recently, new interest in a similar induced coherence setup was revoked by the group of Anton Zeilinger. They showed the possibility of exploiting a change in visibility when scanning a non-uniformly transparent object at the idler path between the two crystals SPDC crystals, which has potential application in quantum imaging \cite{Lemos14}. Although this effect was transferred to a practical application, it was not investigated, up to now, to which degree the phase of the pump laser light is preserved in such an induced-coherence experiment. Therefore we setup an experiment which allowed the variation of the phases of the entangled photons as well as of the pump photons.
As result we could demonstrate how the phase memory of the parametric down conversion controls the induced-coherence and allows the observation of  1$^{st}$-order interference with a single photon as a function of the pump light phase delay. \\

\section{Experimental}
Our experimental setup is shown in Fig. 1. The pump laser (Genesis, Coherent) provides almost diffraction limited cw radiation at 355 nm with a spectral bandwidth of about 45 GHz resulting in a coherence length of about 1.4 mm. A beam splitter was used to divide the pump beam into two almost identical coherent sub-beams to pump two BBO crystals. The crystals serving as parametric down converters had a length of 4 mm and were slightly tilted with respect to the pump beam. The whole setup was arranged with filters and apertures to detect signal photons at a wavelength of 808 nm and the corresponding idler photons at a wavelength of 632 nm.

\begin{figure}[t]
\includegraphics[width=8cm]{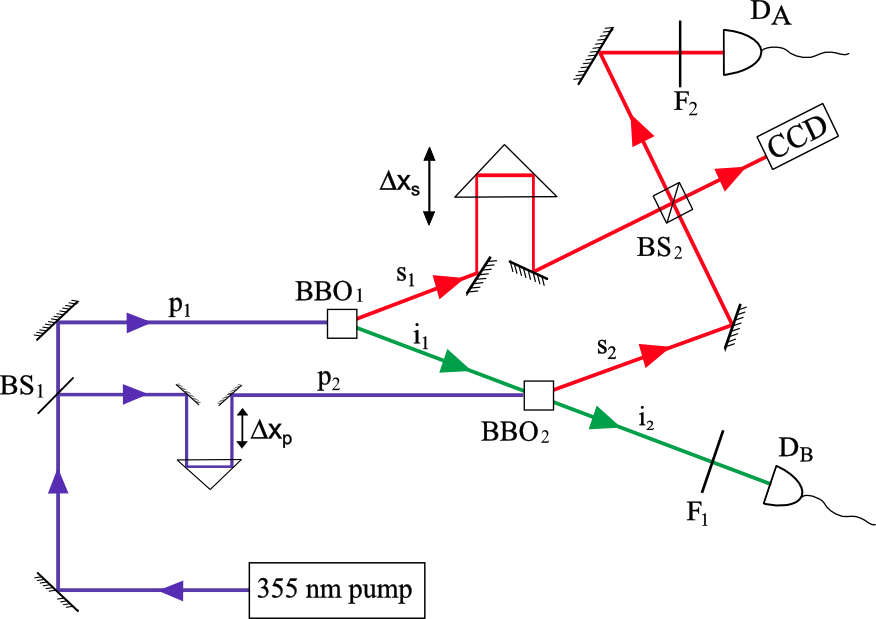}
\caption{Experimental setup for observing phase memory in single photon interference:  For parametric down conversion two 4 mm long BBO crystals were used. The variable pump path length difference $\Delta$x$_{p}$ was realized by a high resolution delay line in the path of pump beam p$_{2}$. The path length difference between the two signal beams $\Delta$x$_{s}$ was realized by a high resolution (minimum step size 10 nm) translation stage. D$_{A}$ and D$_{B}$ are fiber coupled avalanche photo diodes SPCM-AQRH-13 from Perkin Elmer}
\end{figure}

The idler output of the crystal BBO$_{1}$ was used to seed the SPDC process in the crystal BBO$_{2}$ as described in \cite{Zhou91}. Much care had to be taken to match the transversal mode of the idler i$_{1}$ with the one of idler i$_{2}$ generated by the crystal BBO$_{2}$. The signal output of the crystal BBO$_{2}$ was overlapped at the beam splitter BS$_{2}$ with the signal output of crystal BBO$_{1}$. To realize the necessary zero delay between the two signal beam paths at the beam splitter BS$_{2}$ a delay line ($\Delta$x$_{s}$) was inserted in the beam s$_{1}$ between the crystal BBO$_{1}$ and the beam splitter BS$_{2}$. The photons were detected with fiber coupled avalanche photo diodes (SPCM-AQRH-13, Perkin Elmer) at the positions  D$_{A}$ and  D$_{B}$. At  D$_{B}$ both idler photons from the crystal BBO$_{1}$ or BBO$_{2}$ were recorded. With the photo diode  D$_{A}$ the signal photons from crystal BBO$_{1}$ and crystal BBO$_{2}$ were measured. With the filter F$_{1}$ the spectral position around 632 nm and the bandwidth of about 3 nm (FWHM) of the detected idler photons were selected. The filter F$_{2}$ in front of the signal photon detector  D$_{A}$ had a center wavelength of 808 nm and a bandwidth of 2 nm (FWHM). For alignment purposes an EMCCD-camera (Andor Ixus) was positioned at the second port of the beam splitter BS$_{2}$  to check the transversal overlap of the two signal beams from the two crystals.\\
A second delay line $\Delta$x$_{p}$ was positioned in the pump beam p$_{2}$ for the crystal BBO$_{2}$ to vary the phase between the two pump beams in the two crystals. The signal photon interference of 1$^{st}$-order was measured with the detector  D$_{A}$ as a function of either the path length difference of the pump beams $\Delta$x$_{p}$ or the path length difference between the two signal beams $\Delta$x$_{s}$. In addition with the detectors  D$_{A}$ and  D$_{B}$ the coincidence between the signal and the idler photons could be detected as a function of both delays.\\
The pump power for the two crystals was set to 38 mW. The resulting count rates at the detectors were 42,000 photons/sec and 110,000 photons/sec for  D$_{A}$ and  D$_{B}$, respectively. Under these conditions the probability for simultaneous generation of two biphotons during the measuring period of 2 ns was less than 10$^{-2}$. This insured induced-coherence instead of stimulated emission as mechanism for the generation of the second biphoton and thus, the induced emission can be neglected under these circumstances. So far all interference effects between the biphotons generated by crystal BBO$_{1}$ and BBO$_{2}$ were based on phase memory effects.\\

\section {Theory}
To interpret the results a simplified model \cite{Milonni96, Rehacek96} based on field operators is used.  Due to the very low  conversion efficiency $\gamma^2$, we consider strong pump fields p$_{1}$ and p$_{2}$ described by classical complex amplitudes A$_{p1}$ and A$_{p2}$ respectively. For all modes perfect phase matching, monochromatic fields and a point like interaction are assumed. The photon annihilation operators of the the signal and idler modes after crystal BBO$_{1}$ are given by:
\begin{equation}
\begin{aligned}
{a_{s1} = a_{so1} +i ~K_{1} ~ a^{\dagger}_{io1}}\\
{a_{i1} = a_{io1} + i ~K _{1}~ a^{\dagger}_{so1} },
\end{aligned}
\end{equation}
where a$_{so1}$ and a$_{io1}$ are the free, unperturbed operators satisfying the relation $a_{so1}\Ket{ \Psi} = a_{io1}\Ket{ \Psi} = 0$ for the initial state $\Ket{ \Psi}$ with no signal and idler photons. K$_{i}= \gamma A_{pi}$ is a constant under the assumption of no pump depletion.\\
The propagation delay in pump path p$_{2}$ leads to a phase shift $\Delta\varphi _{p}$. This phase shift is accounted for the signal and idler fields after the second crystal BBO$_{2}$:
\begin{equation}
\begin{aligned}
{a_{s2} = a_{so2} + i ~K_{2} ~ e^{i\Delta \varphi _{p}}  ~ a^{\dagger}_{io1}}\\
{a_{i2} = a_{io1} + i ~K_{2}  ~ e^{i\Delta \varphi _{p}}  ~ a^{\dagger}_{so2} + i ~K_{1} ~ a^{\dagger}_{so1} }.
\end{aligned}
\end{equation}
In this equations it is assumed that the idler field i$_{o2}$ that mixes with the pump p$_{2}$ in BBO$_{2}$, to generate the signal  s$_{2}$ field from BBO$_{2}$, is approximately the same as the vacuum idler field i$_{o1}$ that mixes with the pump p$_{1}$ to generate the signal s$_{1}$ in BBO$_{1}$. This assumption is valid as long as the idler fields of both crystals are perfectly aligned and  the conversion efficiency is so low that the generated idler field i$_{1}$ of BB0$_{1}$ contributes negligibly (compared to the vacuum idler) to the generation of the signal s$_{2}$ in the second crystal.
The total signal field a$_{s}$ behind the beam splitter BS$_{2}$ (with transmissivity t and reflectivity r) including the phase shift $\Delta\varphi _{s}$ due to the delay of signal s$_{1}$ is then:
\begin{equation}
\begin{aligned}
a_{s} = & ~t~a_{s2} + r~a_{s1} ~ e^{i\Delta \varphi _{s}}\\
= & ~r a_{so1}e^{i\Delta \varphi _{s}}+ ta_{so2} +i a^{\dagger}_{io1} \big( t K_{2} e^{i \Delta \varphi _{p}} + r K_{1} e^{i\Delta \varphi _{s}} \big) .
\end{aligned}
\end{equation}
Aligning the interferometer for the signal photons s$_{1}$ and s$_{2}$ including the optical path between the crystals BBO$_{1}$ and BBO$_{2}$ via i$_{1}$ results in interference of 1$^{st}$-order for the signal photon at the beam splitter BS$_{2}$. This can be measured using detector D$_{A}$ as shown in Fig. 1. The observed photon expectation rate P$_{A}$ for the initial state $\Ket{ \Psi}$ can be obtained by trivial algebra if equal pump power (K$_{i}=K$) and perfect 50:50 beam splitters are assumed:
\begin{equation}
\begin{aligned}
P _{A}  \approx  \Braket{a^{\dagger}_{s}~a_{s}} \approx  |K|^2   \big(\  1 +  cos( \Delta \varphi _{p} - \Delta \varphi _{s} )   \big) . 
\end{aligned}
\end{equation}
The signal counting rate R$_{A}$ is proportional to the this quantity which exhibits a sinusoidal variation with both phase shifts from the signal or pump fields. Thus the single photon detection contains information not only about the signal photons themselves but also about the pump phase .\\
The phase influence as a function of the delays in the pump beam and in the signal beam can also be observed in the coincidence rate R$_{AB}$ between the detectors  D$_{A}$ and  D$_{B}$:
\begin{equation}
\begin{aligned}
R _{AB} \propto  \Braket{ a^{\dagger}_{s}~a^{\dagger}_{i2} ~ a_{i2}~ a_{s}} 
             \approx |K|^2  \big(\  1 +  cos( \Delta \varphi _{p} - \Delta \varphi _{s} )   \big).
\end{aligned}
\end{equation}
This theoretical description based on a single-mode formalism shows clearly how the interference effect of 1$^{st}$-order measured at D$_{A}$ and the interference effect of the coincidence measured at D$_{A}$ and D$_{B}$ is influenced by both phase differences, between the pump beams on one hand and between the signal beams on the other.\\

\section{Results}
In a first measurement the coincidence rate between the detectors  D$_{A}$ and  D$_{B}$ was obtained as a function of the signal path length difference $\Delta$x$_{s}$ over several periods of the wavelengths. The result is shown in Fig. 2b.

\begin{figure}[t]
\includegraphics[width=8.5 cm ]{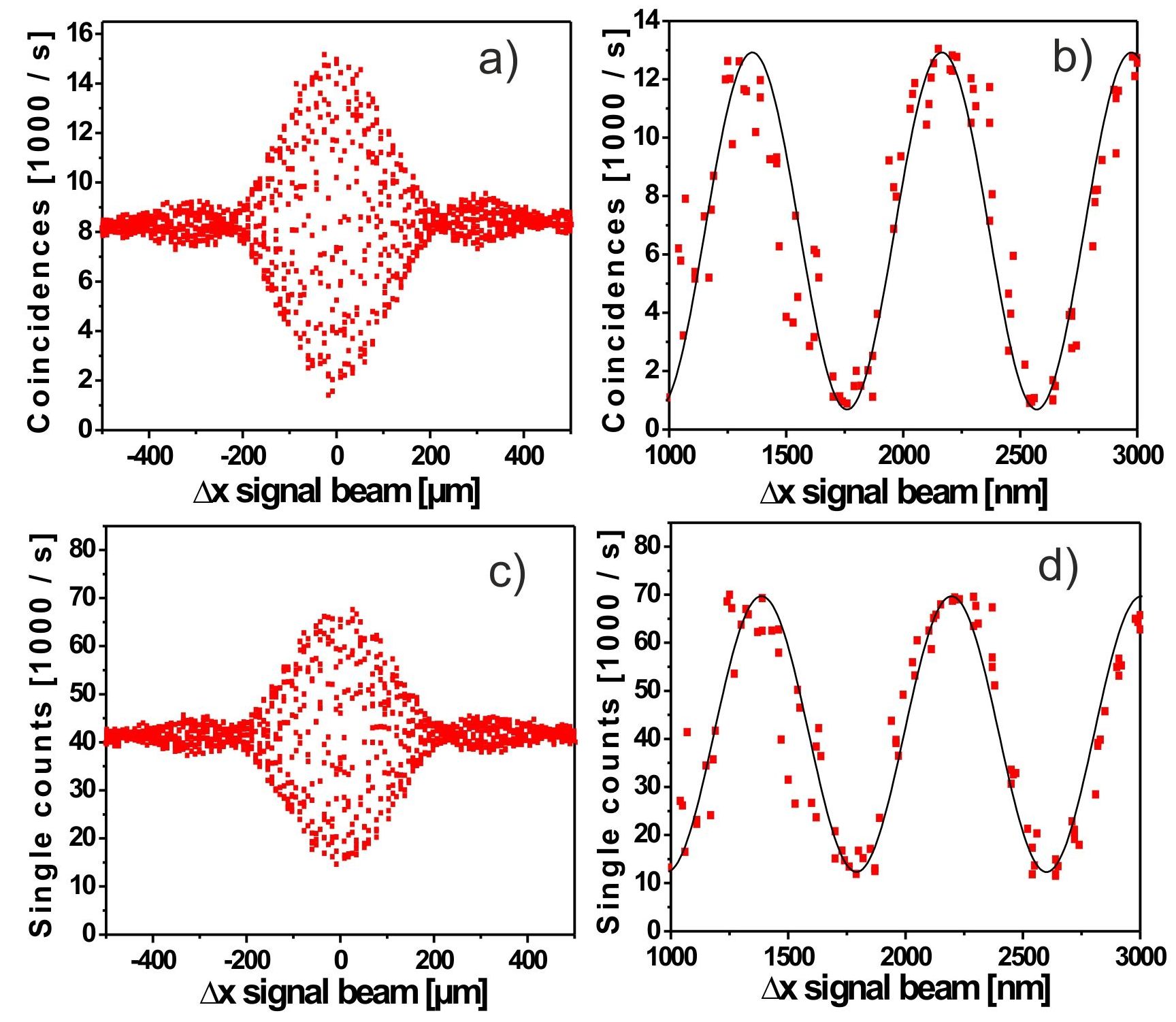}
\caption{ a, b) Measured coincidence rate R$_{AB}$ between detector D$_{A}$ and  D$_{B}$ as a function of the signal path length difference $\Delta$x$_{s}$ and c, d) measured single count rate R$_{A}$ of detector D$_{A}$ as a function of the signal path length difference $\Delta$x$_{s}$  a, c) Coarse scan over the entire region where interference occurs, and b, d) high resolution scan to visualize the fringe period. The solid curves are the best fitting sinusoidal functions resulting in a period of 808 nm.}
\end{figure}

As can be seen in Fig. 2b a good contrast in the coincidence rate could be observed. The visibility of this measurement was approximately 92$\%$. This indicates a high degree of the overlap of the two idler-photon beams towards the detector D$_{B}$. The sinusoidal function used for the best fitting resulted in a period of 808 nm, which corresponds to the wavelength of the signal photon. So far this experimental result is an excellent proof of the induced-coherence effect in the second crystal (BBO$_{2}$).
As discussed above this effect should also be observable in the 1$^{st}$-order interference of the signal photons. The result of the count rate of detector D$_{A}$ alone as a function of the same signal path length difference $\Delta$x$_{s}$ as above is given in Fig. 2d.\\
This 1$^{st}$-order interference shows a modulation with the detuning of the "Mach-Zehnder interferometer"  of the signal photons, as discussed above. The best fitting led again to a period of 808 nm as observed in Fig. 2d. Thus the 1$^{st}$-order interference is in exact agreement with the coincidence measurement . The interference is visible for signal path length differences up to $\pm$200 $\mu$m. The shape and the width of this interference signal is given by the spectral function of the interference filter in front of the detector D$_{A}$. 
The visibility of the 1$^{st}$-order interference reaches a maximum value of 70$\%$ and is lower than in the coincidence measurement. In general the induced coherence setup is very sensitive to the mode matching between the two idler beams. It has been demonstrated by Grayson et al. \cite{Grayson94b} that due to the spatial non-overlap resulting from the divergence properties of the down-conversion the visibility is reduced, even for perfect alignments. They also predict that it is easier to achieve higher visibility in second-order interference than in first order, because the single photon interference is much more critical to any kind of losses between the two idler beams.\\
To demonstrate the phase memory effect in parametric down conversion with induced-coherence over both interferometers we measured the coincidence rate between the detectors  D$_{A}$ and D$_{B}$ but now as a function of the pump path length difference $\Delta$x$_{p}$. During this measurement the path length difference of the signal beams were kept fixed at $\Delta$x$_{s} = 0$. The result is shown in Fig. 3a and 3b. These curves show, to our knowledge for the first time, the influence of the pump phase in an induced-coherence experiment.\\
The visibility in this coincidence measurement is 91$\%$. The best fitting sinusoidal function (solid curve) resulted now in a period of 355 nm, although the interfering signal photons have a wavelength of 808 nm. This period agrees exactly with the wavelength of the pump laser. Regarding the described theoretical background this has to be expected. So far this experiment can be viewed as the biphoton interference effect based on the generation of the biphotons in crystals BBO$_{1}$ or BBO$_{2}$. These two possibilities interfere with each other transporting the phase memory of the generation process of the two biphotons with the pump laser wavelength. So far, as described theoretically, the phase memory of the pump photons is imprinted in the generated pairs of signal and idler photons from the two crystals.\\
But the result of the following measurement seems to be counterintuitive where the same effects are observed in the signal photon count rate of detector D$_{A}$ again as a function of a pump path length difference of $\Delta$x$_{p}$. The result is shown in Fig. 3d.\\
\begin{figure}[t]
\includegraphics[width= 9cm ]{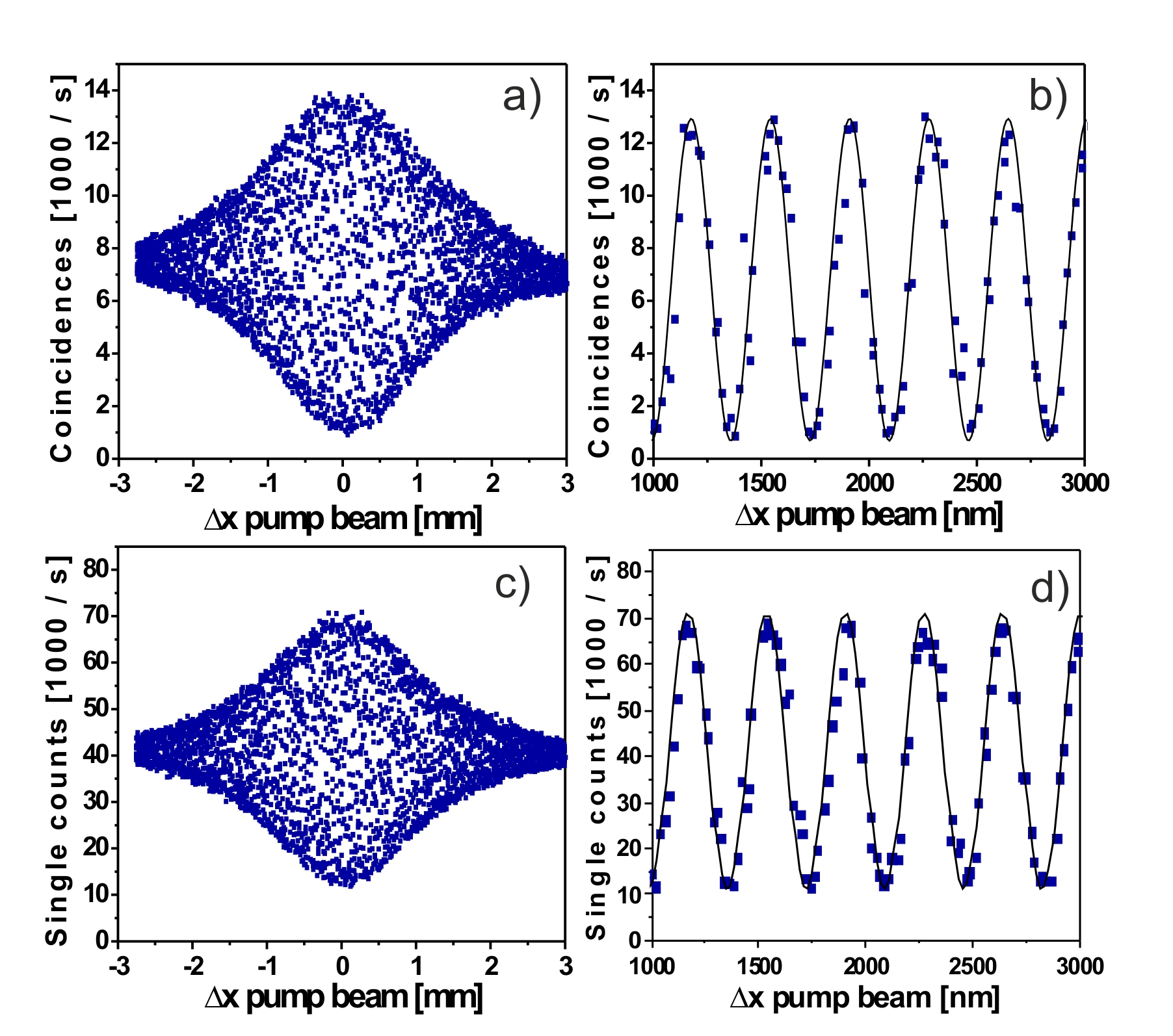}
\caption{ a, b) Measured coincidence count rate R$_{AB}$ between detectors D$_{A}$ and D$_{B}$ as a function of the pump path length difference $\Delta$x$_{p}$ and c, d) measured single count rate R$_{A}$ of detector D$_{A}$ as a function of the pump path length difference $\Delta$x$_{p}$. a, c) Scan over the entire coherence region, and b, d) zoom to visualize the fringe period. The solid curves are the best fitting sinusoidal functions resulting in a period of 355~nm.}
\end{figure}
The best fitting period is again exactly 355 nm as in the previous measurement.  Thus in a single photon interference measurement (via the "Mach-Zehnder interferometer") an interference effect can be observed with about half the wavelength of the observed photons. The pump path difference $\Delta$x$_{p}$ can be detuned over a more than ten times wider range than the signal path difference $\Delta$x$_{s}$ while obtaining interference (see Fig. 3a and 3c). Even for pump photon delays, which exceeds the coherence length of the signal photon by  factor of three the visibility is still  above 90$\%$. This difference of the tuning ranges corresponds to the also about ten times longer coherence lengths of the pump light compared to the coherence length of the signal photons which was given by the interference filter in front of the detector D$_{A}$. As described above, these results can be understood  with a simplified single-mode description. Our experiments demonstrate this quantum effect clearly. The observed interference effect of 1$^{st}$-order using single photons at the beam splitter BS$_{2}$ demonstrates the phase memory of the parametric down conversion in both crystals.\\ 

\section{Conclusion and discussion}
The phase memory effect in SPDC generation was demonstrated by single photon interferometry, to our knowledge, for the first time. The phase memory is initialized by the pump photons in the first crystal and is "transported" via the entangled idler beam to the second crystal. Thus the phase memory could be observed in coincidence. But it was also clearly observed in the interference of 1$^{st}$-order with single photons. In these experiments one part of the "Mach-Zehnder interferometer" was realized as induced-coherence channel from the first to the second crystal. Although the signal photon had a wavelength of 808 nm the modulation of the observed 1$^{st}$-order interference showed a period of 355 nm which is exactly the wavelength of the applied pump laser light.\\
The phase memory is transported across the whole setup. Even the measured single photons in the interference experiment "memorized" the information about the original pump phase relationship. The observed high visibility in these experiments indicates perfect coherence transfer.
The experimental results were theoretically described by a simple model using just one longitudinal mode for the pump, for the signal and for the idler photons, each. So far the experiment demonstrates how phase memory is transported and opens up possible new experiments regarding complementarity and which path information.\\

\section{Acknowledgment}
We are grateful to P. W. Milonni for valuable and illuminating discussions.

\end{document}